\documentstyle[aps,prl,twocolumn,epsf]{revtex}

\begin{document}
\title{Breakdown of Self-Organized Criticality}
\author{Maria de Sousa Vieira\cite{email}}
\address{Department of Biochemistry and Biophysics, University of California,
San Francisco, CA 94143-0448.}
\maketitle
\begin{abstract}
We introduce two sandpile models which show the same behavior 
of real sandpiles, that is, an almost self-organized critical 
behavior for small systems and the dominance of large avalanches as 
the system size increases. The systems become fully self-organized 
critical 
as the system parameters are changed,  showing that these 
systems can make a bridge between the well known theoretical 
and numerical results and what is observed in real experiments.   
A simple mechanism determines the boundary where 
self-organized criticality can or cannot be found, which is the existence 
of local chaos. 
\end{abstract}
\pacs{PACS numbers: 05.45+b, 91.45.Dh.}
\narrowtext
The concept of Self-Organized Criticality (SOC) was introduced by 
Bak, Tang and  
Wiesenfeld (BTW) in 1987 to denote a phenomenon in which 
out of equilibrium, multidimensional systems, 
drive themselves to a critical state characterized 
by a power-law distribution of event sizes\cite{soc}.  
Until then, the studies of fractal structures  
were related to equilibrium systems where this kind of distribution 
appears only at special parameter values where a
phase transition takes place. 
In that pioneering work, the concept of SOC was illustrated by 
a model for sandpiles and since then an  enormous amount of research on SOC, 
both theoretically and experimentally, has been done. Among other 
phenomena in which SOC has been connected with    
are earthquakes\cite{ofc} and evolution\cite{sneppen}.   

The existence of SOC in an experiment with  
a quasi-one-dimensional pile of rice
was demonstrated by 
Frette {\sl et. al}\cite{frette}.   
They found that 
the occurrence of SOC depends on the shape of the rice. 
Only with sufficient 
elongated grains avalanches with a power-law distribution 
occurred. For more symmetric grains a stretched exponential distribution 
was seen. 
Christensen {\sl et. al}\cite{oslo} 
introduced 
a model for the elongated rice pile experiment in which the local 
critical slope varies 
randomly between 1 and 2. They found that their model, 
known as the  Oslo rice pile model, reproduced 
well the experimental results on the quasi-one-dimensional rice pile. 
In a recent publication\cite{cmltrain}, we introduced a 
fully deterministic one-dimensional SOC system,  
which presents the same qualitative and 
quantitative behavior of the Oslo system. In other words,   they  
belong to the  same universality 
class. 

When one goes to sandpiles  
with geometry of two-dimensions a 
different picture emerges. That is, the  models\cite{soc} 
predict the 
presence of power-law distributions and the experiments do not 
display them. The most well known sandpile experiments can be 
classified in two types: 
(a) local dropping of sand in the center of the pile\cite{held} and  
(b) uniform driving, more specifically  
the rotating drum experiments\cite{nagel,nori}. 
In class (a) it was found  that small systems 
present scaling almost consistent with SOC, but in large systems another 
regime with big avalanches belonging to a different distribution 
appears.   
In class (b)  
one sees small avalanches  
that seems to display a power-law  distribution of limited size. 
These small avalanches are interwoven with big system-wide 
avalanches which belong to a different distribution.   
There are studies of avalanches in natural settings\cite{himalaya}, which  
also show similar distribution as the ones observed in the sandpile 
experiments.

Although SOC has not been observed in sandpile experiments
is well known that power law distributions  does exist in Nature, 
one of the most well known cases being Gutenberg-Richter law 
for earthquakes\cite{gutenberg} in agreement with       
SOC. It is clear that 
there is a missing piece in this puzzle. That is, would there 
be  simple models that would display the observations in real sandpiles, 
and still present SOC in other parameter regions?  
To our knowledge it still missing in the literature such    
models and it is the aim of this letter to introduce them.  
Our models makes a bridge between the SOC and non SOC behavior, and 
the boundary that characterizes the separation between the 
two regimes is the absence or presence of local chaos. 
However, we will see that  local 
chaos is a necessary but not sufficient condition for SOC to be 
destroyed.  
  
Our models are inspired by the  model introduced 
in Ref.\cite{cmltrain}, and is also  
governed by coupled lattice map
(that is, systems characterized by discrete time and continuous 
values for the space variables).  
Here we increase the dimensionality and change 
the drive and the relaxation rules.   
Having in mind the sandpiles 
experiments we first introduce a model for the  
local dropping of sand. We assume a two-dimensional square lattice 
of linear size $L$   and  
to each site $i,j$ in the lattice there is  
associated to it     
a variable $x_{i,j}$ with $x \in [0 , +\infty)$, which is to 
represent  the local slope of the pile.    
The dynamics of the model is described by the following algorithm:

(1) Start the system by assigning random initial values for the variables 
$x_{i,j}$,   
so the they are 
below a chosen, fixed,  threshold $x_{th}$.     

(2) Choose a nearly central site  of the lattice  
and update it slope  
according to $x_{i,j}=x_{th}$. 

(3) Check the slope in each element. If an element $i,j$ has $x_{i,j} \ge x_{th}$, 
update  $x_{i,j}$ according to $x'_{i,j}=\phi(x_{i,j}-x_{th})$, where 
$\phi $ is a given nonlinear function that has two parameters 
$a$ and $d$. 
Increase the slope in all its nearest neighboring element according 
to  $x'_{nn}=x_{nn}+\Delta x/4$, where 
$\Delta x=x_{i,j}-x_{i,j}'$ and $nn$ denotes nearest neighbors.   

(4) If $x'_{i,j} < x_{th}$ for all the elements, go to step (2) (the event, or 
avalanche,  
has finished). Otherwise, go 
to step (3) (the event is still evolving).    

Without losing generality, we can take $x_{th}=1$. 
In our simulation in step (2) we have chosen the site with 
$i=j=L/2$.
The nonlinear function we use is 
\begin{equation}
\phi (x) = \cases{ 1-d-ax, \ \ \ \  {\rm if} \ \ x < (1-d)/a,\cr 
           0, \ \ \ \  {\rm otherwise.}\cr} 
\label{eq1}
\end{equation}
The parameter $d$ would be associated with the minimum drop in energy
after an event involving one single element and $a$ would be the parameter 
associated with the amount of friction between the grains. That is,  
the smaller the $a$, the larger the friction and 
the smaller the change in the slope of the pile.   
We have tested several other functions  
and found that the quantitative and qualitative results 
we show here are robust. The important 
ingredient being the shape of $\phi (x) $ in the vicinity of $x=0$.  
In contrast with the one-dimensional case\cite{cmltrain}, 
it is not required here  
that $\phi (x) $ be periodic in order to find the presence SOC. 

We have chosen to evolve the system using parallel dynamics with 
open boundary conditions. It is beyond the scope of the present letter 
to study the several possible variations of our models. Further results 
on these models will be presented in a future publication\cite{future}. 
The distribution of time duration of the avalanches will also be 
presented in the future, but our preliminary results show that they 
are qualitatively similar to the ones for the size distribution.   

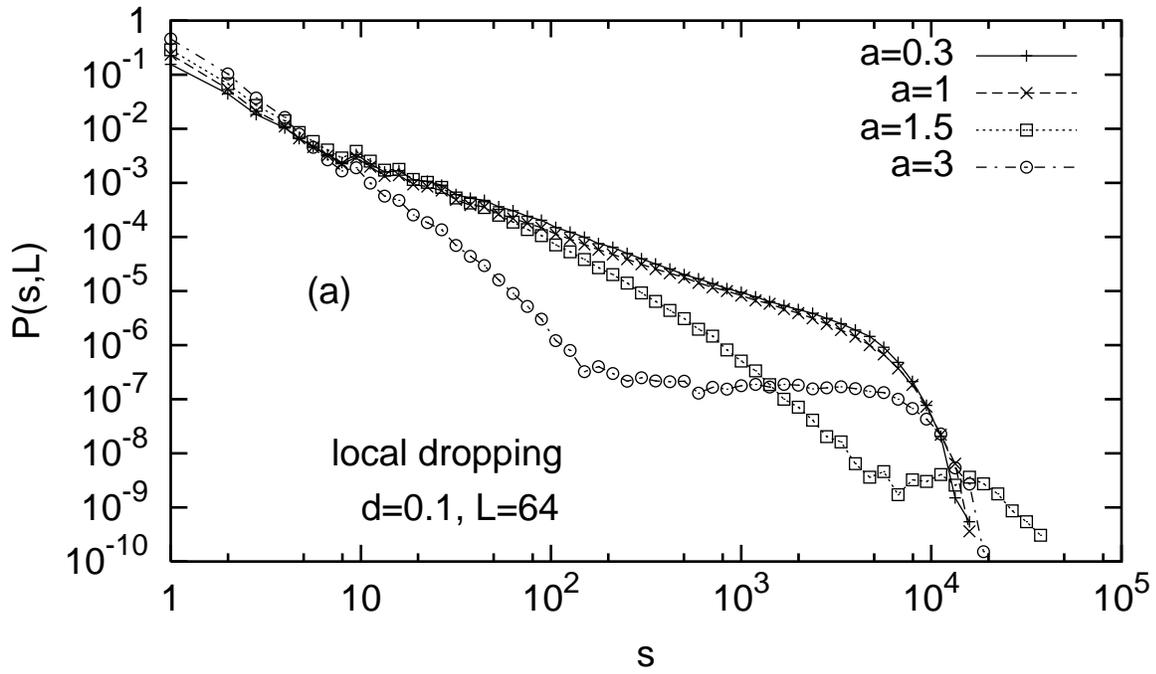
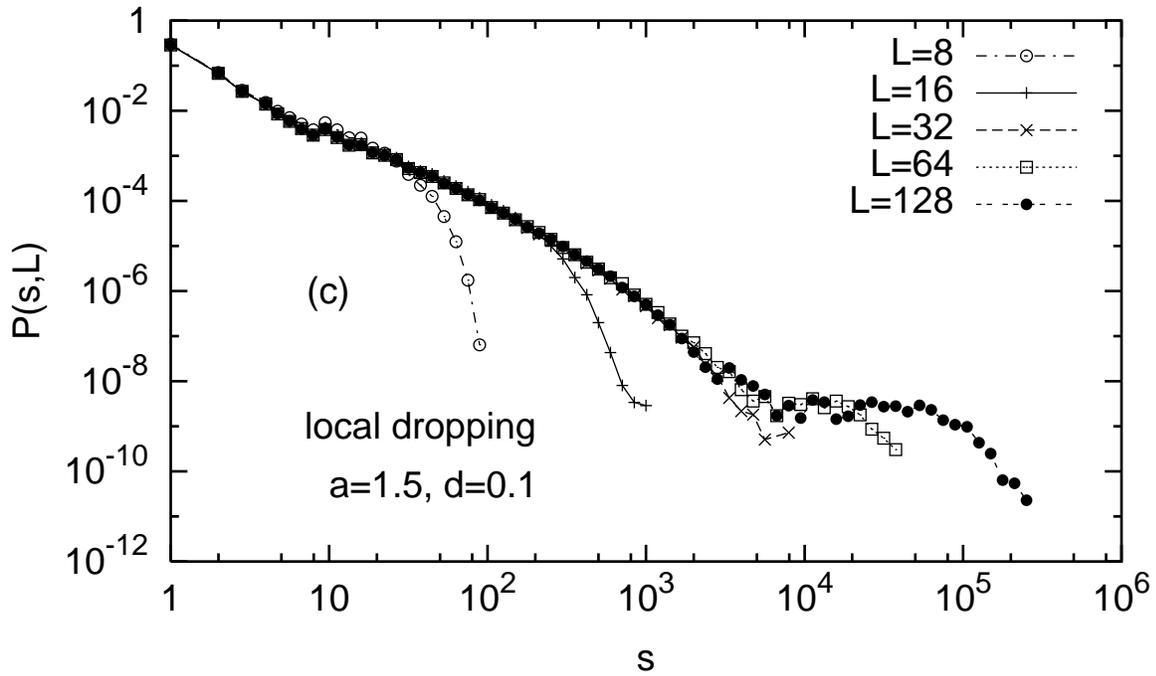
\begin{figure}
\caption[f1]{
(a) Frequency of events involving $s$ updates for (a) variable $a$ with 
$d = 0.1$, and (b) variable $d$ and $a=1.5$ with 
$L=64$. In (c)  we vary the system size and use $a=1.5$ and $d=0.1$, and 
in (d) we show the fitting using the scaling of Eq. 2 for small 
systems, with $\beta =6$ and 
$\nu =3 $. In this figure we are using the model for local dropping of 
sand. We have used $4 \times 10^6$ avalanches in all the simulations of this 
paper.   
}
\label{f1}
\end{figure}

We 
display an example of our simulations in 
Fig.~\ref{f1}, where  we show the distribution of events $P(s)$ involving 
$s$ update steps, that is, the size of the avalanche.  
The events that involve all the 
elements of the system  have been excluded from our analysis. However, 
in this model of local dropping of sand  we have observed 
that nothing very distinct will occur if they are also included in 
the statistics.   
In (a) we show $P(s)$ for  
$L=64$,  $d=0.1$ and 
vary $a $ and in 
(b), we use $a = 1.5 $ and vary $d $, keeping $L=64$. 
We notice the existence of two regimes. 
For small $d$, power-law distributions, that is SOC, 
appears only when $a \le 1$, 
whereas if $d$ is large, we see SOC even 
with $a > 1$. 
As in real sandpiles\cite{held}, 
the almost SOC regime is 
characterized  by 
a  region with an apparent scaling for small events, and 
the big events  
belong to a different distribution. To illustrate this, we show 
in Fig.~\ref{f1}(c)  simulations  with $a = 1.5$ and $d=0.1$ and 
varying $L$. 
In small systems  $P(s)$ can be fit to 
a scaling form of the type  
\begin{equation}
P(s,L)=L^{-\beta}G(s/L^\nu),    
\label{eq2}
\end{equation}
as shown in 
Fig.~\ref{f1}(d), where we have used $\beta = 6$ and $\nu = 3$. 
The function $G$ is not 
well fit by a power-law, since one can clearly see in the figure that it 
is curved. We have found\cite{future} that it is 
consistent with a stretched exponential, as in real sandpiles\cite{stretched}.
The observations of 
Himalayan avalanches\cite{himalaya} also display the kind of distribution 
shown in Fig.~\ref{f1}(c) for large systems.        

The second model we introduce here is for the rotating drum 
experiment\cite{nagel,nori}, which we call uniform drive, 
since the slope of the pile increases uniformly for all the grains.  
The algorithm is similar to the one described above, with the 
exception of step (2), which is now replaced by 

(2) Find the element in the lattice that has the largest $x$, 
denoted here by $x_{max}$.  
Then update all the lattice  elements according to    
$x_{i,j}=x_{i,j}+x_{th}-x_{max}$. 

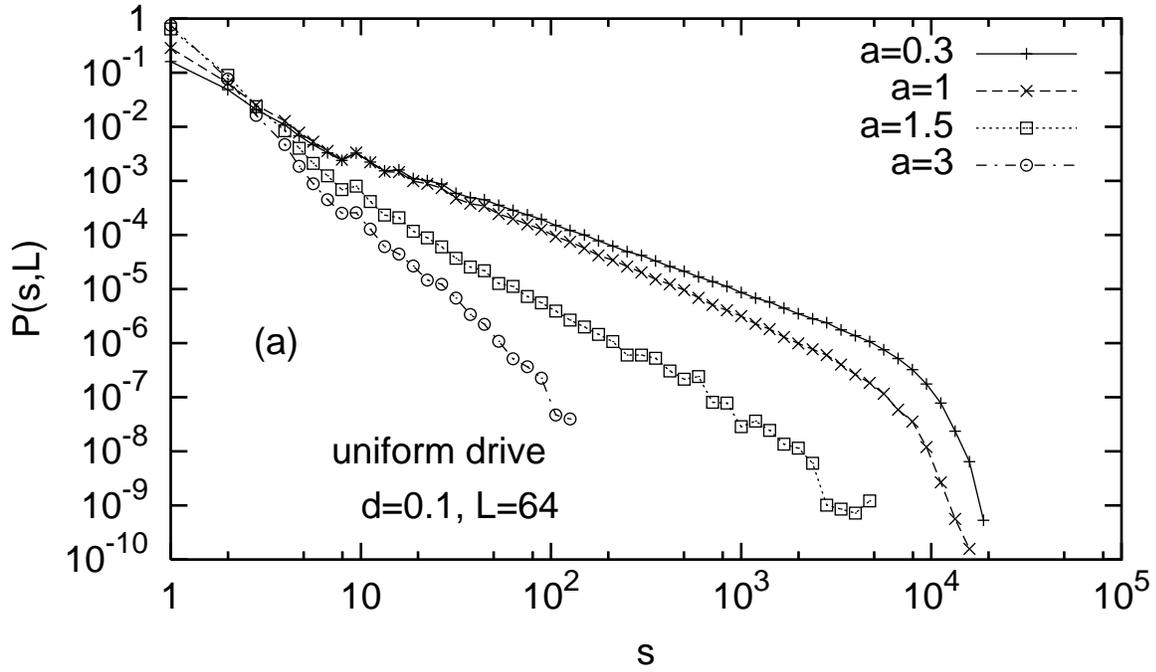
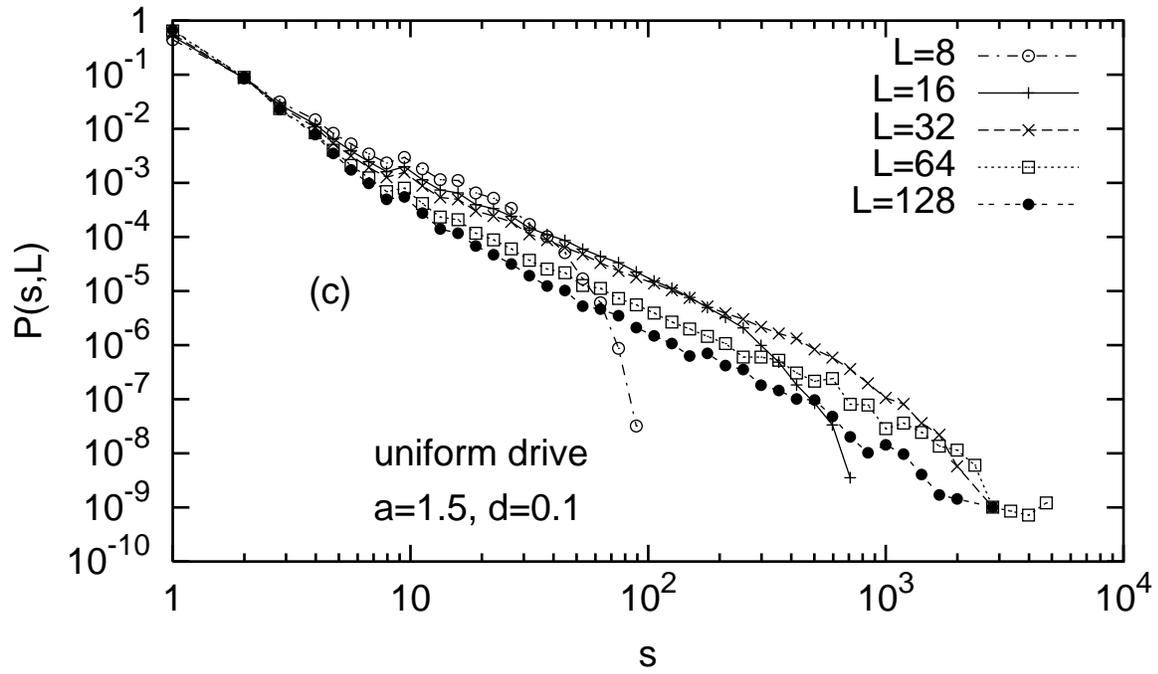
\begin{figure}
\caption[f2]{
(a) Frequency of events involving $s$ updates for (a) variable $a$ with
$d = 0.1$, and (b) variable $d$ and $a=1.5$ with 
$L=64$. In (c)  we vary the system size and use $a=1.5$ and $d=0.1$. 
In this figure we are using the model for uniform drive (that 
is, the rotating drum experiment).  
}
\label{f2}
\end{figure}

We show examples of $P(s)$ for this model in Fig.~\ref{f2}.
There, in (a) we fix $d$ and vary $a$, and in 
(b) we fix $a$ and vary $d$. In both cases we have used $L=64$ and 
the events that involve all the elements of the system have been excluded.  
Distinctly from the model of local dropping, it seems here that 
there is a power-law distribution for any parameter value. However, 
the behavior is not exactly SOC.  We have found  that SOC is only seen if 
$a \le 1$ or $d $ is large enough, as in the case of the local 
dropping. When $a > 1$ and $d$ is smaller than a given value, 
we see a SOC-like behavior only  for small values of $L$. As $L$ grows, 
there is a transition to a different behavior, in which the 
larger the system, the smaller the  power-law region, as 
Fig.~\ref{f2}(c) shows. 
 
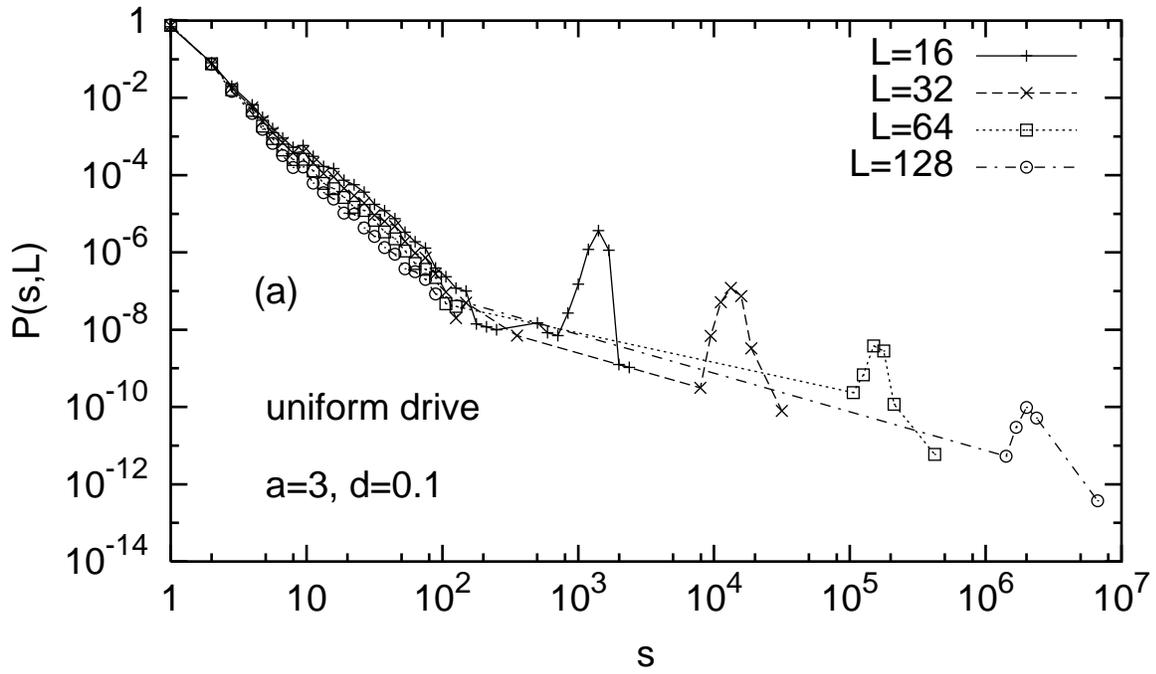
\begin{figure}
\caption[f3]{
(a) Frequency of events involving $s$ updates for 
$a=3$ and  $d=0.1$ with varying $L$. The peaks in the distribution 
are the events involving all the elements of the system. 
In (b) we show the fitting using the scaling of Eq. 2 for the region of 
avalanches of intermediate sizes, using  
$\beta =1$ and
$\nu =0 $. 
In this figure we are using the model for uniform drive. 
}
\label{f3}
\end{figure}

System-wide avalanches have been reported in the rotating drum 
experiment\cite{nagel,nori} that belong to a different distribution
than the one of the small avalanches. This is exactly      
what we see in this model for $a > 1$ and small $d$. 
In Fig.~\ref{f3}(a) we show all the  events  of the system
including the ones that involve all the elements  
(for $a=3$, $d=0.1$ and varying $L$). 
We see that the scaling region does not get bigger as the 
system size increases, and a peak related to the events involving 
all the elements is seen. We have found that the intermediate size events 
($10 \lesssim s \lesssim 100$) 
can be fit by a scaling of the type 
$P(s,L)=L^{-\beta}G(s/L^\nu)$ with $\beta = 1$ and $\nu = 0$, as 
shown in Fig.~\ref{f3}(b). As in the rotating drum 
experiment\cite{nori,stretched} the function $G$ in this case 
is closer to a power-law than in the case of the experiment with local dropping 
of sand.  

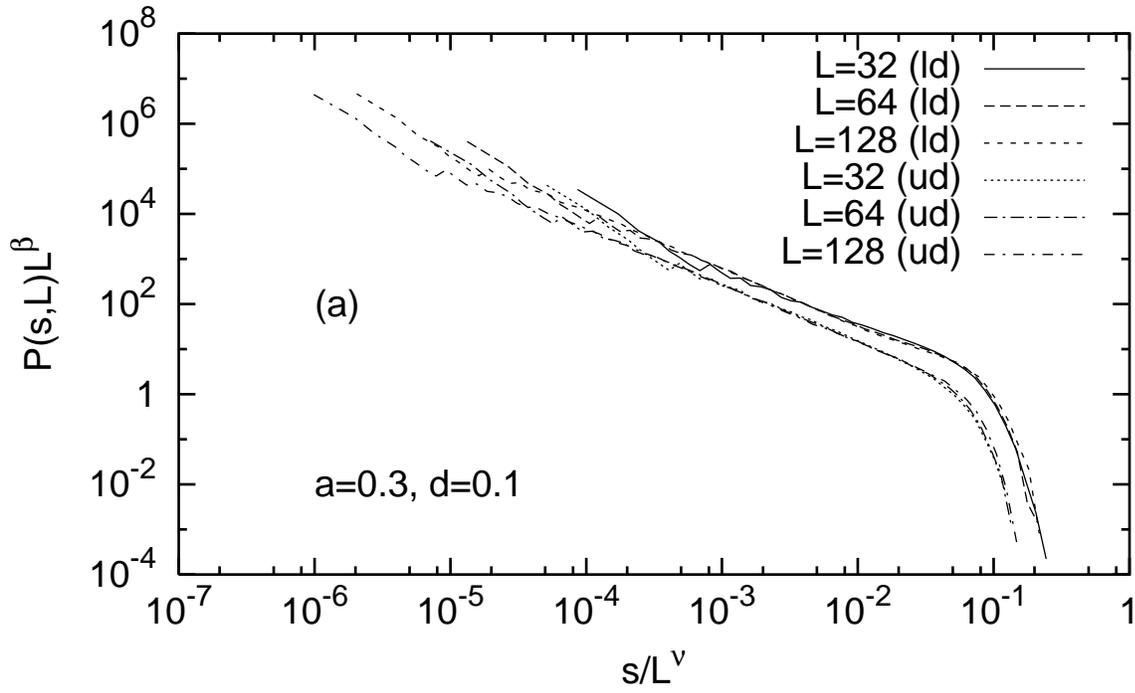
\begin{figure}
\caption[f4]{
(a) Frequency of events involving $s$ updates for
$a=0.3$ and  $d=0.1$ with varying $L$ using the scaling of Eq. 2. 
For the case of local dropping we have used $\beta = 3.55$ and $\nu =2.75$. 
For uniform drive we have used $\beta = 3.55$ and $\nu =2.85$. 
In (b) we show the cases  in which after relaxation  the variable 
$x$ is a given random number uniformly distributed between $0$ and 
$0.99$ for local dropping of sand (ld) and for uniform drive (ud). 
We also show the cases in which $x$ is relaxed to 0 after an event, which 
is the same as having $d=1$. 
}
\label{f4}
\end{figure}

For given $a$ and $d$ 
the slope of the power-law distribution  seems to be 
the same for both models, that is, local dropping of sand 
and uniform drive, but the slope varies  with $a$, as shown in Fig.~\ref{f1} and 
~\ref{f2}. Consequently, the universality class of these models 
vary with the parameters.  
In Fig.~\ref{f4}(a) we show 
the scaling given by Eq.~(\ref{eq2}) with $\beta = 3.55$ and 
$\nu = 2.70 $ for the local dropping (ld) and $\beta = 3.55$  $\nu = 2.85$ for 
the uniform drive (ud). 
In the limit $a \to 0$ or $d \to 1$ the slope of 
the power-law distribution is the same as the one in the BTW model\cite{soc} 
and the conservative OFC model\cite{ofc},  
that is, $P(s) \sim s^{-1.245}$.  

We next investigate what would happen if the relaxation function 
is just a random number generator.   
In other words, instead of using  $\phi $ in
step  3 of the above algorithm we now use
$x'_{i,j}=\rho $, where $\rho $ is a random number uniformly 
distributed in the interval $[0, 1-d]$. We have found SOC  
for any  $d \in (0, 1]$ with the same  
exponents as the BTW model. This  
is displayed in Fig.~\ref{f4}(b)  where we show the size distribution 
for $d=0.01$ in the cases of local dropping (ld) and uniform 
drive (ld). Therefore nonlinearities in $\phi $ and consequently 
non ergodicity are necessary for the SOC behavior to be destroyed 
in these models.  
In that figure, we also show the case in which 
$x'_{i,j}=0$, which corresponds to $d=1$. In this limit we recover 
the OFC\cite{ofc} model for the case of uniform drive.

The reason why $a=1$ determines a special boundary, in which 
SOC may or may not be present, is due to the fact that it 
marks the boundary in which local chaos exists. 
By ``local" here we mean that is is at the grain level, no matter what happens 
at the system level. Chaos 
is defined as the exponential divergence of trajectories that 
start with almost the same initial conditions. In fact, if we 
consider two copies of a given system, copy 1 and copy 2, with all 
the elements having the same $x$ except that in copy 1  the element 
$x^1_{i,j}=x^*$, whereas in copy 2 $x^2_{i,j}=x^*+\Delta$.   
It is not difficult to find that after one iteration by $\phi $, the 
separation of the two elements instead of $\Delta $ will be $a \Delta $. 
So, if $a>1$ the separation increases (that is, we have local chaos) 
whereas if $a < 1$ the separation decreases.  We have seen from the 
above results that, since SOC can exist even with $a > 1$, if $d$ is 
large enough or $L $ is smaller than a given value,   
this means that local chaos is a necessary but 
not sufficient condition for SOC to be broken. Consequently, the 
true boundary between SOC and non SOC depends on three parameters, 
that is, $a$, $d$ and $L$. Further studies are necessary to 
understand this boundary more clearly. 
   
In summary, we have introduced a model for sandpiles, or other  
systems that present avalanche like behavior, that reproduce very 
well the observation in real sandpiles. We have found  that a simple 
mechanism, i.e local chaos,  
can explain the breakdown of SOC  in those systems.  
Based on our results, we believe that with an appropriate choice 
of grains SOC will be seen even in real sandpiles. Grains with 
large friction would be the best candidates for this.

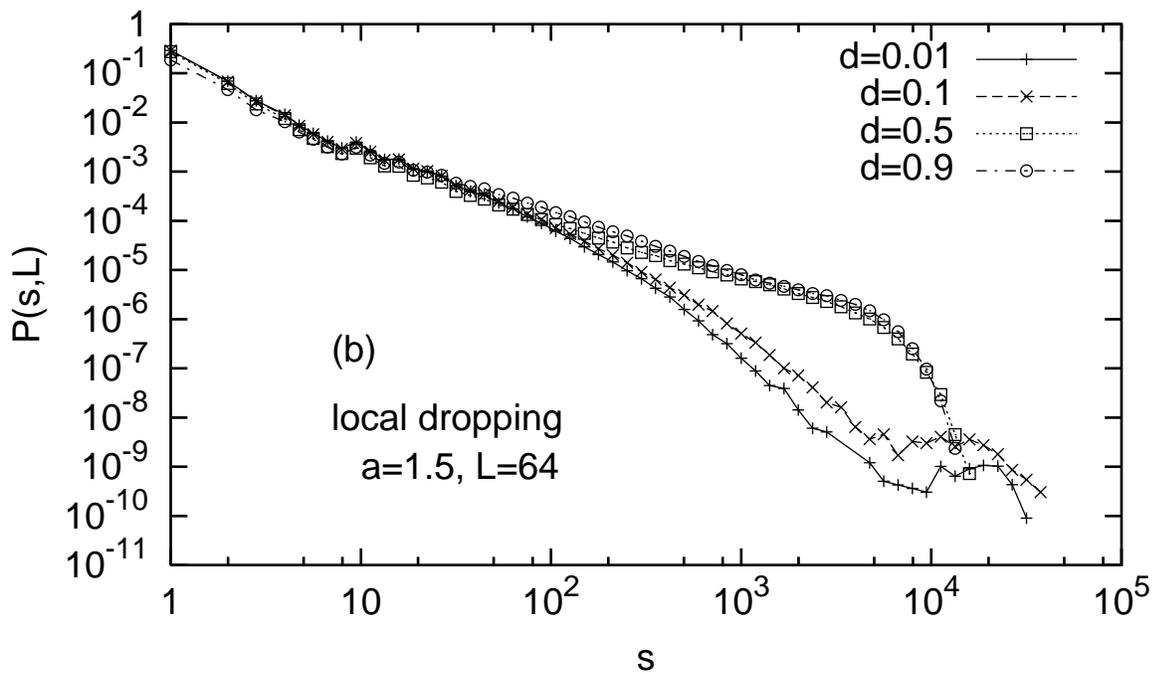
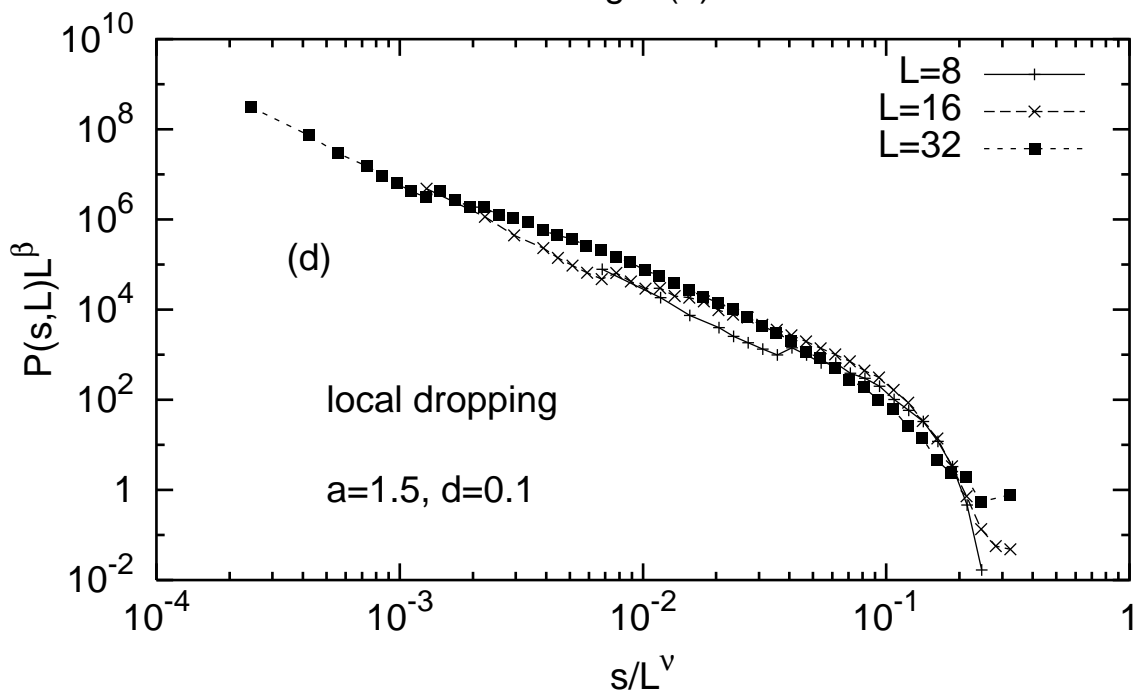
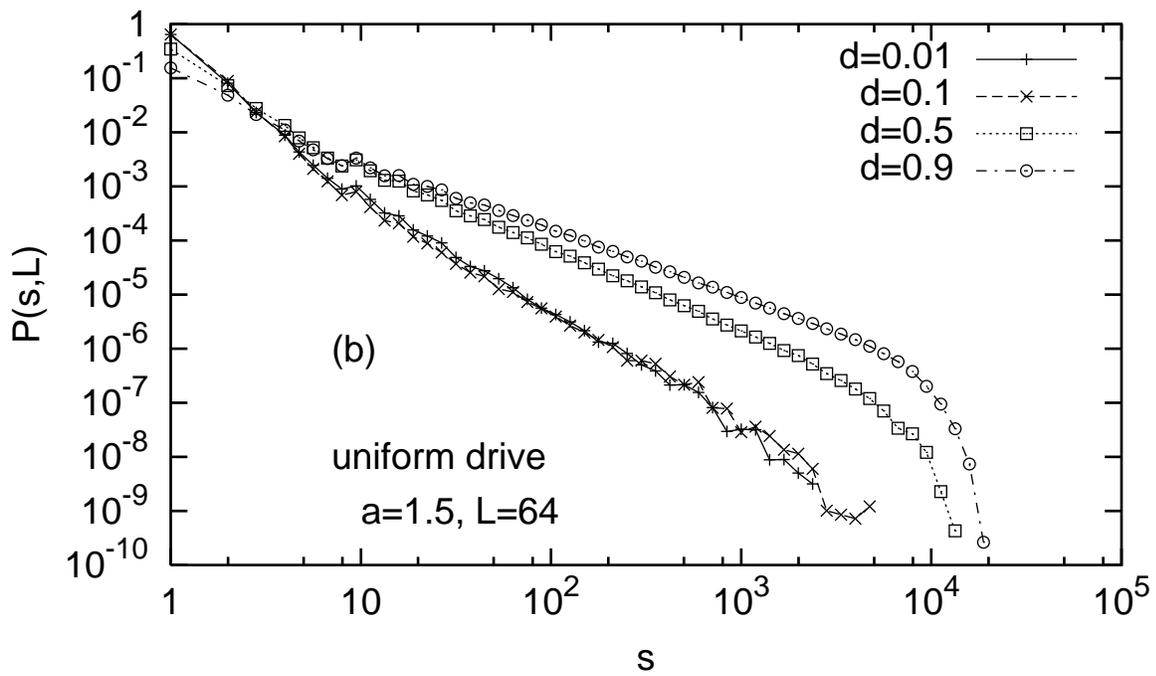
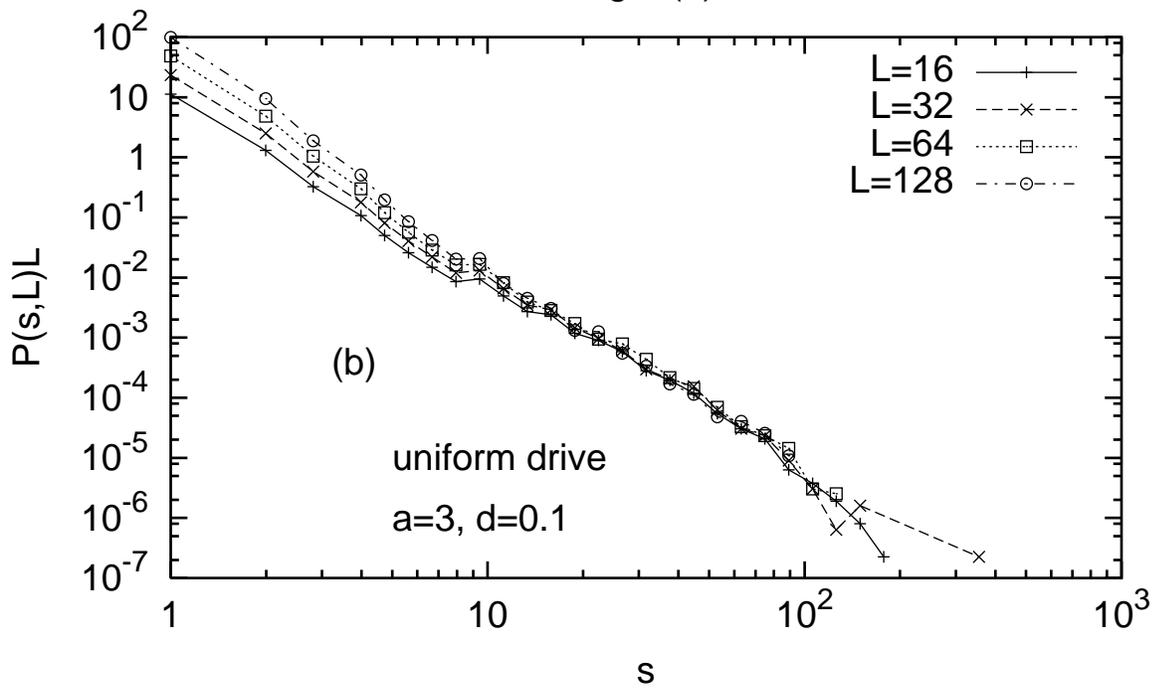
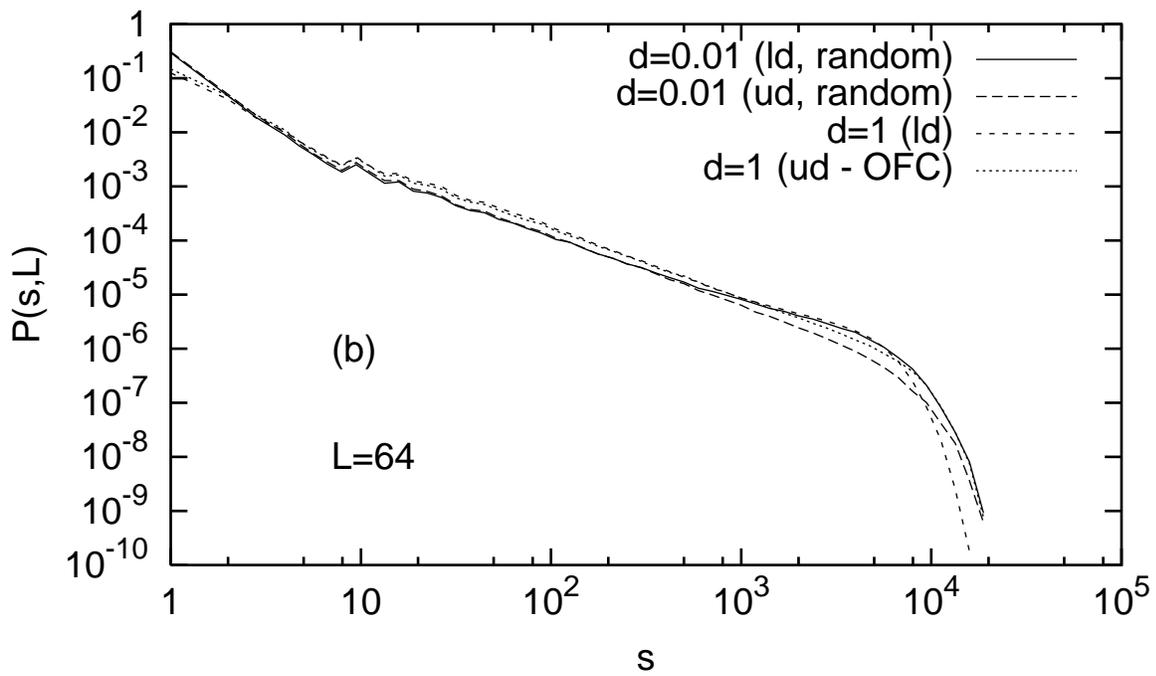
\end{document}